\documentclass[12pt]{article}

\usepackage{scicite}

\usepackage{times}

\usepackage{graphicx}

\usepackage{multirow}

\usepackage{xcolor}

\usepackage[pagewise]{lineno}

\newenvironment{sciabstract}{%
\begin{quote} \bf}
{\end{quote}}

\title{Andromeda's asymmetric satellite system as a challenge to cold dark matter cosmology}

\author
{Kosuke Jamie Kanehisa,$^{1,2}$ Marcel S. Pawlowski,$^{1}$ Noam Libeskind$^{1}$\\
\\
\normalsize{$^{1}$Leibniz-Institut für Astrophysik Potsdam; An der Sternwarte 16, 14482 Potsdam, Germany}\\
\normalsize{$^{2}$Institut für Physik und Astronomie, Universität Potsdam;}\\
\normalsize{Karl-Liebknecht-Straße 24/25, 14476 Potsdam, Germany}\\
}

\date{}

\begin{document} 


\baselineskip24pt


\maketitle


\begin{sciabstract}
    The Andromeda galaxy is surrounded by a strikingly asymmetrical distribution of satellite dwarf galaxies aligned towards the Milky Way. 
    The standard model of cosmology predicts that most satellite galaxy systems are near-isotropic, and dwarf associations observed in the local Universe are only weakly asymmetric. 
    Here, we characterise the Andromeda system's asymmetry, and test its agreement with expectations from concordance cosmology. 
    All but one of Andromeda's 37 satellite galaxies are contained within 107 degrees of our Galaxy. 
    In standard cosmological simulations, less than 0.3\% (0.5\% when accounting for possible observational incompleteness) of Andromeda-like systems demonstrate a comparably significant asymmetry. 
    None are as collectively lopsided as the observed satellite configuration. 
    In conjunction with its satellite plane, our results paint the Andromeda system as an extreme outlier in the prevailing cosmological paradigm, further challenging our understanding of structure formation at small scales. 
\end{sciabstract}

\emph{Published in Nature Astronomy, DOI:10.1038/s41550-025-02480-3}


\newpage
\section*{The asymmetric M31 satellite system}

Dwarf satellite galaxies serve as the building blocks of structure formation in the standard Lambda cold dark matter ($\Lambda$CDM) paradigm of cosmology \cite{Sales2022baryonic}, and are predicted to be distributed in a near-isotropic manner around their more massive host galaxies \cite{Kroupa2005great}.
The unexpected presence of flattened and co-rotating planes of satellites around the Milky Way \cite{Pawlowski2012vpos,Pawlowski2020milky} and Andromeda \cite{Ibata2013vast} galaxies, as well as a handful of dwarf associations beyond the Local Group \cite{Muller2021coherent,Chiboucas2013confirmation,Muller2017m101,Martinez-Delgado2021tracing,Ibata2015eppur,Heesters2021flattened,Pawlowski2024satellite}, are hence difficult to explain in the context of concordance cosmology.
Similarly anisotropic satellite distributions are highly rare ($<1\%$) in cosmological simulations based on the $\Lambda$CDM framework \cite{Pawlowski2020milky,Ibata2014thousand,Muller2021coherent}, sparking an active and enduring debate on whether this discrepancy indeed constitutes a challenge to our understanding of small-scale structure formation \cite{Bahl2014comparison,Cautun2015planes,Samuel2021planes,Sawala2023milky}.

While satellite planes are generally characterised under an assumption of axial symmetry, an alternate signature of spatial anisotropy can be found in the strikingly asymmetrical distribution of satellites around the Andromeda galaxy (M31).
Around 80\% of M31's satellites lie within a hemispheric region facing the Milky Way \cite{Savino2022hubble}, a fraction that has persisted throughout improvements in dwarf distance estimates and the discovery of new satellites over the past two decades \cite{McConnachie2006satellite,Conn2013three-dimensional}.
While the excess of satellites detected on the near side of Andromeda may hint at the role of photometric completeness, the distribution of satellites brighter than the limiting magnitude $M_V\leq-7.5$ at the far side of M31's virial radius \cite{Doliva-Dolinsky2022pandas} demonstrates a similar degree of lopsidedness to that of the full sample \cite{Savino2022hubble}.
Surveys with a given footprint also cover a larger physical volume on the far side of Andromeda, counteracting the comparative ease of detecting satellites at closer distances.
The observed asymmetry is likely of physical origin, although its formation history remains unclear \cite{Wan2020origin}.

While the Andromeda system hosts the most populated and well-studied dwarf association that demonstrates a prominent asymmetry in its spatial distribution, a weaker degree of lopsidedness among satellites of paired host galaxies appears to be prevalent in the local Universe.
One group \cite{Libeskind2016lopsided} examined tens of thousands of Local Group-like host galaxy pairs in the Sloan Digital Sky Survey (SDSS), detecting up to a 10\% excess of satellites in the region between paired hosts.
A comparable overabundance was later found for analogous host pairs in cosmological N-body simulations \cite{Pawlowski2017lopsidedness}.
Curiously, satellites observed around isolated host galaxies also demonstrate a tendency to lie on the same side of their hosts \cite{Brainerd2020lopsided,Brainerd2021mean} to a degree roughly consistent with their simulated analogs \cite{Wang2021lopsided}.

Nevertheless, these large-volume studies are limited by their low median sample size of only 3-4 satellites per host.
These small samples hinder any in-depth analysis of their individual satellite distributions, especially at the far end of the satellite luminosity function.
This motivates a closer examination of the M31 system, which remains the most populated, strongly asymmetric satellite association in our cosmic neighbourhood.
Savino \emph{et al.} \cite{Savino2022hubble} recently released a set of RR Lyrae-based distance measurements for 39 stellar systems in the M31 system.
This sample includes nearly all known satellites within Andromeda's virial radius of 266 kpc \cite{Fardal2013inferring} and several others, as well as the Andromeda galaxy itself to anchor them.
Adopting these homogeneous distances, we investigate the nature of Andromeda's asymmetric satellite distribution and quantify the incidence of similarly lopsided analogs in $\Lambda$CDM cosmological simulations.

We use a combination of HST archival data and newly obtained distances compiled in Savino \emph{et al.} \cite{Savino2022hubble}.
Of the 39 stellar systems listed, we mirror the authors' analysis by excluding the Giant Stellar Stream.
Accordingly, we make use of heliocentric distances for 37 satellite galaxies and M31 itself.
This sample includes most satellites located within the PAndAS survey footprint (out to a projected $\sim 150\,\mathrm{kpc}$ from M31), but contains several additional dwarfs at larger radii (And XXXI -- XXXIII) \cite{Martin2013lacerta,Martin2013perseus}.
Of the 37 M31 satellites, 7 lie beyond Andromeda's adopted virial radius.
For ease of comparison with previous work on the Andromeda system \cite{McConnachie2006satellite,Conn2013three-dimensional,Savino2022hubble}, satellite positions are transformed to a Cartesian reference frame aligned with the Galactic coordinate system, centred upon the expected position of M31 as shown in Fig.~\ref{fig:1}.
Andromeda's galactic disk is aligned with the longitudinal plane and the positive z-axis ($b=90^{\circ}$) points to M31's north galactic pole, while the Milky Way lies at the same azimuth as the positive x-axis ($l=0^{\circ}$).
Distance uncertainties are accounted for by generating 10,000 Monte Carlo realisations per satellite.

The on-sky distribution of Andromeda's satellites as seen from their host galaxy is plotted in Fig.~\ref{fig:2}.
Even upon a cursory examination, the area-preserving property of the adopted Aitoff-Hammer projection highlights a remarkable dearth of satellites on the far side of M31 -- only a single bright dwarf, NGC 205, lies within $73^{\circ}$ of the line-of-sight.
29 satellites ($78\%$) are contained within the hemisphere facing the Milky Way, while up to 32 satellites (86\%) lie on one side of the Andromeda galaxy -- the normal vector to the plane of maximum asymmetry separating them with the remaining 5 satellites is oriented to within $17^{\circ}$ of the Milky Way's direction.
The geometric centroid of Andromeda's satellite distribution is displaced by as much as $75\pm15\,\mathrm{kpc}$ from M31's position and lies at a $37^{\circ}$ offset from our Galaxy (reduced to $22^{\circ}$ when disregarding the outlying Peg DIG and IC 1613).
The M31 satellites' asymmetric distribution is evidently aligned quite closely with the Milky Way.

The commonly used, hemisphere-based metric for quantifying the M31 system's lopsidedness  \cite{McConnachie2006satellite,Conn2013three-dimensional,Savino2022hubble,Doliva-Dolinsky2022pandas} struggles to identify less-populated satellite agglomerations constrained to small solid angles or "voids" in angular space unnaturally devoid of satellites.
For improved flexibility, we adopt an alternative method as follows.
We generate conical regions with opening angles within $\theta=[0.5,180]^{\circ}$ in $0.5^{\circ}$ increments (where $\theta=180^{\circ}$ corresponds to a full sphere), oriented along $10^5$ isotropically distributed vectors on a unit sphere.
For each $\theta$, we identify the corresponding cone which encloses the maximal population $N(\theta)$ of Andromeda's satellites (Fig.~\ref{fig:3}), which we refer to as the \emph{cone of maximum asymmetry}.
To quantify the M31 system's lopsidedness specifically towards the Milky Way, we additionally generate one \emph{companion-locked} cone per $\theta$ oriented in our Galaxy's direction.
The latter approach is equivalent to the hemisphere-based metric at $\theta=90^{\circ}$.
In Fig.~\ref{fig:2}, cones of maximum asymmetry with lower $\theta$ are concentrated in the southwestern quadrant along smaller clusters of satellites, many of which participate in M31's satellite plane \cite{Ibata2013vast,Savino2022hubble}. At larger opening angles (and correspondingly larger $N(\theta)$ in Fig.~\ref{fig:3}), cone orientations demonstrate a markedly closer alignment with the Milky Way.

To determine the statistical significance of the observed asymmetry, we compare the M31 satellites with an isotropic satellite distribution.
If every galaxy has a $50\%$ chance to be on either side of M31, the probability of finding at least 29 out of 37 satellites on the near side of Andromeda is $4\times10^{-4}$.
The hemispheric M31 satellite distribution is therefore statistically different from the isotropic model with a significance of $3.3\sigma$.
We can also extend this analysis to opening angles other than $180^{\circ}$.
Each companion-locked cone covers a fraction $\sin^{2}(\theta/2)$ of a unit sphere's surface.
The most significant configuration of Andromeda's satellites lies within $\theta=106.5^{\circ}$ with a corresponding binomial probability of $1.6\times10^{-6}$.
The full Andromeda system is distinct from an isotropic distribution of satellites at the $4.7\sigma$ confidence level.

\section*{Comparison to $\Lambda$CDM simulations}

Although the observed satellite distribution is unlikely to occur by chance, this alone does not allow us to determine the Andromeda system's agreement with expectations from $\Lambda$CDM cosmology.
Indeed, a weak degree of asymmetry appears to be prevalent among satellite galaxy systems in cosmological simulations \cite{Pawlowski2017lopsidedness,Wang2021lopsided}.
This anisotropy can be attributed to the accretion of subhaloes from preferential directions, along cosmic filaments and in gravitationally bound groups \cite{Zentner2005anisotropic} -- thus forming a significant population of dynamically young satellites that have yet to relax in their host galaxy's halo \cite{Gong2019origin}.
To determine whether these processes are sufficient to explain the observed asymmetry in Andromeda's satellite distribution, we examined the incidence of similarly lopsided distributions in two cosmological simulations: TNG100 \cite{Springel2018results} and EAGLE \cite{McAlpine2016eagle}.
We further validate the robustness of our results in the higher-resolution TNG50 run \cite{methods}.
In addition to gravitational effects and structure formation in a standard Planck cosmology, both simulation suites include baryonic processes such as stellar feedback, and the disruption of dwarf satellites due to enhanced tidal forces from the host galaxy's baryonic disc \cite{Garrison-Kimmel2017not}.

We identify M31 analogs by finding dark matter haloes with masses $M_{200}$ between $5\times10^{11}\,M_{\odot}$ and $3\times10^{12}\,M_{\odot}$, a range centred upon but broadened from the mass criteria previously used in studies of M31's satellite plane \cite{Bahl2014comparison,Ibata2014thousand} for improved statistics.
For each simulated host galaxy, we search for nearby haloes with mass $>0.25\,M_{200}$ within $5\,R_{200}$ \cite{methods} and classify systems as paired or isolated by whether a companion is found.
Each M31 analog is required have at least 37 satellites within $2\,R_{200}$, an initial sample from which the 37 most massive satellites are selected.
We fix the satellite population of simulated systems since metrics of phase-space correlation demonstrate an intrinsic correlation \cite{Pawlowski2017considerations} with the number of satellites.
We thus obtain 1268 and 1107 systems from TNG100 and EAGLE respectively, of which 184 and 141 have a companion galaxy.

Distance uncertainties are traditionally accounted for by sampling the prior of the observed satellites.
However, this approach is equivalent to applying uncertainties twice -- once upon observation, and once more when compensating for the initial distance errors.
If Andromeda's true satellite distribution were heavily lopsided, we would therefore systematically wash out the existing asymmetry.
We instead mock-observe the simulated host galaxies and their satellite distributions at M31's distance of $776^{+22}_{-21}\,\mathrm{kpc}$ \cite{Savino2022hubble}, applying the observed satellite distance uncertainties to generate 100 MC realisations per simulated system \cite{methods}.

Fig.~\ref{fig:3}A shows the population of satellites contained within cones of maximum asymmetry around Andromeda's simulated analogs.
Satellite systems in a $\Lambda$CDM context demonstrate a considerable degree of lopsidedness.
All 37 satellites are contained within $\theta=135^{\circ}$ (or equivalently, there exists an angular void of $\theta=45^{\circ}$) around over $70\%$ of M31 analogs.
Nevertheless, the observed Andromeda system deviates significantly from the simulated median at larger opening angles.
This deviation peaks at a cone of maximum asymmetry with $\theta=101^{\circ}$ containing 36 out of 37 satellites, aligned to within $6^{\circ}$ of the Milky Way's direction.
We find that the incidence of a similarly asymmetric configuration is $0.45\%$ in both TNG and EAGLE (Fig.~\ref{fig:4}A).
The departure from $\Lambda$CDM expectations is enhanced when solely considering asymmetry oriented towards a companion galaxy in Fig.~\ref{fig:3}B, which peaks at a companion-locked cone with $\theta=106.5^{\circ}$ also containing all but one satellite.
In simulations, none of TNG and EAGLE's 325 paired M31 analogs can reproduce the Andromeda system's observed asymmetry towards our Galaxy (Fig.~\ref{fig:4}B), effectively setting an upper limit of $0.3\%$ on the frequency of M31-like lopsided systems.

The Andromeda system demonstrates a degree of asymmetry that, in its current configuration, is remarkably rare among its simulated analogs.
And yet, it may be argued that our selected $\theta$ was specifically chosen to maximise Andromeda's discrepancy with simulations -- we do not know a priori the opening angle at which simulated systems are most significantly lopsided.
To compare the M31 system's generalised asymmetry to $\Lambda$CDM expectations, we need to take into account the \emph{look-elsewhere effect}, a statistical phenomenon wherein an apparently significant result may arise by chance due to the extended parameter space searched \cite{Cautun2015planes}.

To compare Andromeda and its analogs on a balanced playing field, we generate a reference sample of $10^6$ systems, each consisting of 37 unit vectors drawn from isotropy.
For each simulated system (and the observed Andromeda system itself), we obtain the fraction $f_{\mathrm{iso}}(\theta)$ of this reference sample that matches or exceeds the given system's satellite population $N(\theta)$ within its cone of maximum asymmetry.
A companion vector is also randomly drawn to enable a similar analysis using companion-locked cones instead.
For each family of cones, we determine the system's minimum $f_{\mathrm{iso}}(\theta)$ across all opening angles, which we refer to as its \emph{minimum isotropic frequency} (MIF).
By adopting the MIF in lieu of $N(\theta)$ as our metric of lopsidedness, we compare satellite systems in their most significantly asymmetric configurations.

Using cones of maximum asymmetry, as many as $4.5\%$ of M31 analogs in TNG and $6.2\%$ in EAGLE match Andromeda's MIF of $7\times10^{-5}$.
Paired and isolated analogs demonstrate similar incidences at $5.6\%$ and $5.3\%$ respectively.
Although the significance of a majority of these systems is driven by smaller clusters of satellites rather than a collective asymmetry \cite{methods}, we recover a tenfold increase in incidence over the fixed-$\theta$ comparison in Fig.~\ref{fig:4}.
In contrast, Andromeda's satellite distribution remains an outlying case with regards to its asymmetry towards the Milky Way.
Only $0.3\%$ of TNG and EAGLE realisations can reproduce the observed MIF of $2\times10^{-6}$, thus yielding an incidence that remains similar to the previous fixed-$\theta$ comparison.
Strikingly, we find that none of these few realisations are driven in significance by a companion-locked asymmetry at an opening angle as large as M31's $106.5^{\circ}$ or a dwarf population above 33 satellites (Fig.~\ref{fig:5}).
The Andromeda system's asymmetry towards our Galaxy occurs at a more population-wide scale than any of its similarly significantly lopsided analogs in cosmological simulations.
Repeating this method for the observationally complete sample of 30 M31 satellites \cite{Savino2022hubble} instead yields an incidence of 0.5\%, which serves as an upper bound to the simulated frequency of M31-like asymmetric systems \cite{methods}.

\section*{Implications for structure formation}

We have shown that Andromeda's asymmetric distribution of satellite galaxies towards the Milky Way is highly unexpected in a full cosmological context.
The formation of this anisotropic structure around M31 poses a puzzle that -- given its rarity among simulated analogs -- requires a unique evolutionary history to explain.
Given the excellent alignment of the Andromeda system's asymmetry with our Galaxy, could the Milky Way potential play a role in maintaining the observed lopsidedness?
If that were the case, assuming that the two Local Group host galaxies are of similar mass, the Milky Way satellites should also experience a comparable effect from M31's potential.
And yet, while difficult to reliably ascertain due to uneven sky coverage due to survey footprints and obscuration from the Galactic disc, no significant degree of asymmetry has been reported in the Milky Way's satellite distribution \cite{Pawlowski2021phase-space}.
Tidal effects are also unlikely to be strong enough to reproduce the observed asymmetry \cite{Bowden2014asymmetric}.
Nevertheless, the $6^{\circ}$ alignment between M31's most significant cone of maximum asymmetry and the Milky Way's direction occurs only at a $\mathrm{sin}^2(\theta/2)=0.27\%$ likelihood if the orientation of the observed asymmetry is random -- strongly hinting at the role of our Galaxy as significant to the formation or evolution of the lopsided Andromeda system.

One group \cite{Wan2020origin} explored the possibility of a single accretion of a well-populated association of satellites forming the current Andromeda system.
Given the observed distribution's wide radial distribution and range of orbital energies, however, they found that the resulting asymmetric structure would likely dissolve as soon as $500\,\mathrm{Myr}$ later.
In this scenario, the prominent asymmetry observed in the Andromeda system must be dominated by a dynamically young population of satellites.
A similar excess of recently accreted satellites may also exist around the Milky Way \cite{Hammer2021gaia,Taibi2024portrait}.
While the weak lopsided signal in simulated paired hosts is indeed driven by satellites recently accreted from nearby filaments \cite{Gong2019origin}, such processes are already self-consistently included in the cosmological simulations searched.
Furthermore, if the observed asymmetry were driven by recent-infall satellites, the well-populated nature of M31's lopsided satellite distribution would imply a catastrophic dearth of satellites with an infall time older than the asymmetry's dispersion timescale.

Unlike the apparent ubiquity of correlated planes of satellites in our cosmic neighbourhood \cite{Pawlowski2020milky,Ibata2013vast,Muller2021coherent,Chiboucas2013confirmation,Muller2017m101,Martinez-Delgado2021tracing,Ibata2015eppur,Heesters2021flattened,Pawlowski2024satellite}, statistical samples of dwarf associations in the local Universe \cite{Libeskind2016lopsided,Brainerd2020lopsided,Samuels2023lopsided} generally demonstrate a degree of lopsidedness consistent with simulations.
This strengthens the need for a unique evolutionary history for M31 within a $\Lambda$CDM context.
Nevertheless, the limited satellite populations in these samples may mask individual systems that are comparably asymmetric as the Andromeda group.
Next-generation surveys with lower surface brightness limits \cite{Habas2020newly,Carlsten2022exploration,Geha2017saga} will be necessary to conclusively determine whether the incidence of individual, highly asymmetric satellite systems also aligns with cosmological expectations.

In this Research Article, we have demonstrated that the satellite galaxies around Andromeda form a strongly asymmetric distribution aligned with the Milky Way.
Even when accounting for the look-elsewhere effect, similarly lopsided configurations of satellites only occur around $<0.3\%$ of M31 analogs in $\Lambda$CDM cosmological simulations, and no single simulated system can simultaneously reproduce the population-wide nature of the observed lopsidedness.
At present, no known formation mechanism can explain the collective asymmetry of the Andromeda system \cite{Wan2020origin}.
In conjunction with M31's plane of satellites, which holds a similar degree of tension with simulations \cite{Ibata2014thousand}, our results present the satellite galaxy system of Andromeda as a striking outlier from expectations in $\Lambda$CDM cosmology.

\section*{Methods}

\renewcommand{\thefigure}{S\arabic{figure}}
\renewcommand{\thetable}{S\arabic{table}}

We make use of RR Lyrae-based distances for the Andromeda galaxy and 37 of its dwarf satellites compiled in \emph{Savino et al.} \cite{Savino2022hubble}.
From their published sample of stellar systems, we exclude the heavily disrupted Giant Stellar Stream (GSS).
We include IC 1613 and the Pegasus Dwarf (Peg DIG) in our working sample despite their outlying distance from M31, but confirm that their exclusion does not significantly affect our results.
The recently discovered Peg V \cite{Collins2022pegasus} lacks the deep HST imaging necessary for the RR Lyrae approach.
Its associated TRGB distance of $692^{+33}_{-31}\,\mathrm{kpc}$ places it firmly on the near side of M31, intensifying the existing asymmetrical trend.
Dwarf coordinates are taken from the Local Volume catalog \cite{Karachentsev2004catalog,Karachentsev2013updated}, a compilation of galactic objects within 10 Mpc from the Local Group.

We account for the associated distance uncertainties by generating 10,000 Monte Carlo realisations as follows.
For each realisation, distance errors are drawn from a Gaussian distribution for each satellite galaxy (with a standard deviation equivalent to the quoted uncertainty) and applied to their expected positions along the respective line-of-sight vectors.
A distance error for M31 is then drawn in the same way, and is further applied to all satellite positions along M31's line-of-sight -- thus yielding realisations still centred upon the system's host galaxy.
Note that errors modelled by this Monte Carlo approach are only used to quantify the properties of the observed M31 system (and hence in Fig.~\ref{fig:2}).

\subsection*{Primary metrics of asymmetry}

The most straightforward manner in which one may quantify the degree of asymmetry in a given satellite system is by counting the number of satellites on each side of an arbitrarily oriented plane, which effectively divides the system into two hemispheres. 
By repeating this check for 100,000 planes with isotropically distributed normal vectors, we find the plane which demonstrates the greatest disparity between the two hemispheric satellite populations -- we refer to this plane as the "plane of maximum asymmetry".
The two satellite populations on either side of the plane of maximum asymmetry is usually expressed as a ratio, but since all sampled systems have exactly 37 satellite galaxies, we simply quote the number of satellites in the more populated hemisphere as the system's maximum hemisphere population $N_{\mathrm{lop}}$.
The orientation of the system's asymmetry is expressed as the normal vector of the plane of maximum asymmetry, pointing towards the direction of satellite excess.
It is important to note that more than one hemisphere containing the same, maximised set of satellites can exist if the maximum angular range spanned by the satellites is marginally less than $180^{\circ}$.
Hence, the orientation of a system's lopsidedness according to the hemisphere metric entails a small degree of intrinsic error.

For systems of paired hosts, an often-used variation of this metric \cite{Conn2013three-dimensional,Pawlowski2017lopsidedness, Savino2022hubble} effectively "locks" the dividing plane's normal vector to the vector pointing from the satellites' host galaxy to its companion: the "pair vector", $\hat{x}_{\mathrm{pair}}$.
The population of the hemisphere oriented in the same sense as the pair vector is denoted $N_{\mathrm{lop,pair}}$, although it and $N_{\mathrm{lop}}$ are frequently used interchangeably in the literature due to the M31 system's hemispheric asymmetry being roughly oriented towards the Milky Way. 

While the most intuitive metric by far, the maximum hemisphere population of a given satellite distribution fail in identifying satellite clusters constrained to a small solid angle or angular regions unnaturally devoid of satellites.
For a more comprehensive characterisation of satellite excesses and underdensities, we define a series of cones with corresponding opening angles $\theta$ (up to a full sphere at $\theta=180^{\circ}$).
For each cone, we generate 100,000 isotropically distributed so-called "grid" vectors, centre the cone on each grid vector, and identify the number of satellites enclosed within the conical volume.
For each $\theta$, we record the most populated cone -- with an associated population $N_{\theta}$ -- and its orientation.
As such, each of these \emph{cones of maximum asymmetry} may be oriented in different directions with varying $\theta$.
We generate 359 cones per system, spanning a range of $\theta=[0.5,180]^{\circ}$ in 0.5-degree increments.
Much like the hemisphere metric, the orientation of the cones may be "locked" to a host galaxy's pair vector $\hat{x}_{\mathrm{pair}}$ \cite{Pawlowski2017lopsidedness}.
In this case, the number of satellites in each \emph{companion-locked cone} with opening angle $\theta$ is denoted $N_{\theta,\mathrm{pair}}$.

A weakness of both of the above approaches is the unintuitive nature of deriving $N_{\mathrm{lop}}$ and $N_{\theta}$ expected in an isotropic satellite distribution.
In the limit of an infinitely populated system, we can approximate the satellite distribution as a constant-density field, and may hence expect the hemisphere or cone of maximum asymmetry to enclose $1/2$ or $\Omega/4\pi$ of the satellite population (where $\Omega=4\pi\sin^{2}(\theta/2)$ is the solid angle associated with the cone's opening angle $\theta$).
However, with a realistic and limited number of satellites, there will generally exist a hemisphere or cone that manages to capture more than the above expectation.
A system of two satellites will always yield $N_{\mathrm{lop}}=2$, for instance, unless the satellites happen to be perfectly diametrically opposed.
Conversely, the companion-locked variants follow the naive expectation from isotropy if pair vectors are randomly drawn.

\subsection*{Supplementary metrics of asymmetry}

Lopsidedness in satellite galaxy distributions is a nuanced topic, and the metric adopted to quantify it will determine what aspects of the asymmetry are most represented. While we adopt the cone-based approach as our primary metric of asymmetry in this work, we also consider several supplementary metrics previously used in other work.

The hemisphere and cone approaches discussed above solely focus on the angular distribution of satellites while disregarding their distance from their host. Hence, the metrics are less sensitive to lopsidedness caused by a translation of the satellite distribution away from their host. To account for this, we calculate the unweighted centroid of satellites for a given system and measure its distance from the host galaxy's position. Since this centroid shift is determined by both the system's degree of lopsidedness and the characteristic distance of its subhaloes, we normalise the shift by $d_{\mathrm{rms}}$, the root-mean-square distance of the satellites from their host. This yields $d_{\mathrm{norm}}$, the normalised centroid shift from the system's host galaxy. It should be noted that $d_{\mathrm{norm}}$ is highly sensitive to outlying satellites, and conversely insensitive towards those that are centrally distributed within their system. One group \cite{McConnachie2006satellite} previously reported a three-dimensional centroid shift of $84\,\mathrm{kpc}$ derived for 16 of M31's satellites, which drops to $50\,\mathrm{kpc}$ when removing the outlying Peg DIG and IC 1613 from their sample. Using the larger satellite sample from \emph{Savino et al.} \cite{Savino2022hubble}, we recover a similar centroid shift of $75\pm15$ kpc (or $62\pm17$ when excluding the two outlying satellites).

One of the most common metrics of lopsidedness found in the literature is the distribution of pairwise satellite angles \cite{Conn2013three-dimensional,Wang2021lopsided}. In a given satellite system, each unique combination of two satellites has a corresponding angular separation, $\theta_{ij}$. The mean of $\theta_{ij}$ across all possible satellite pairs -- the mean pairwise angle $\theta_{\mathrm{lop}}$ -- should tend towards $90^{\circ}$ in an isotropic distribution. A heavily lopsided distribution implies that a majority of satellites occupy a small region of angular space, and hence a reduced $\theta_{\mathrm{lop}}$.  This approach can also be adapted for host galaxies with associated companions. Instead of measuring pairwise angles between all satellites, many studies \cite{Libeskind2016lopsided,Pawlowski2017lopsidedness,Gong2019origin,Brainerd2020lopsided} instead measure the distribution of angles, $\theta_{i,\mathrm{pair}}$, between each satellite and the system's $\hat{x}_{\mathrm{pair}}$. The corresponding mean angle is denoted $\theta_{\mathrm{lop,pair}}$. The same expectation of $90^{\circ}$ holds for isotropic systems, while $\theta_{\mathrm{lop,pair}}<90^{\circ}$ and $\theta_{\mathrm{lop,pair}}>90^{\circ}$ indicates the satellites are lopsided towards and away from the companion galaxy respectively. The M31 system demonstrates a mean pairwise angle of $\theta_{\mathrm{lop}}=83^{\circ}\pm6^{\circ}$ and a mean alignment with the Milky Way of $\theta_{\mathrm{lop,pair}}=70^{\circ}\pm10^{\circ}$ (when sampling distance errors from the Andromeda satellites).

Finally, the Mean Resultant Length (MRL) is a circular statistic that effectively describes the directionality of a satellite distribution, and was recently adopted in \cite{Brainerd2021mean} as an alternative to their previous, pairwise angle approach \cite{Brainerd2020lopsided}. For a system consisting of $N$ satellites, each with position $\mathbf{x}_i$ relative to their host, the MRL is defined by summing satellite unit vectors and dividing the resulting vector's magnitude by $N$:

\begin{equation}
\bar{R} = \left| \frac{1}{N} \sum_{i=1}^N \frac{\mathbf{x}_i}{\left|\mathbf{x}_i\right|} \right|.
\end{equation}
\label{eq:mrl}

A perfectly isotropic distribution should yield $\bar{R}=0$, whereas an unrealistic system with satellites located along a single angular position would produce $\bar{R}=1$. The Andromeda system yields a MRL vector of magnitude $R=0.36^{+0.10}_{-0.14}$ oriented at $28^{\circ}$ from the line-of-sight to the Milky Way (see Fig.~\ref{fig:2}).

\subsection*{Sampling M31 analogs from simulations}

We use data from the IllustrisTNG suite of large-volume cosmological simulations \cite{Springel2018results}, publicly available at \texttt{www.tng-project.org}.
All runs contained therein adopt cosmological parameters taken from Planck \cite{PlanckCollaboration2016}: $\Omega_{\Lambda} = 0.6911$, $\Omega_\mathrm{m}=0.3089$, and $h=0.6774$.
This suite includes three hydrodynamic runs with varying volume and resolution, as well as their dark matter-only counterparts.
We elect to use the intermediate-volume TNG100-1 run due to the relatively large sample of M31 satellites which needs to be matched by any simulated analogs.
TNG100-1 spans a simulation box of length $110.7\,\mathrm{Mpc}$ with dark and gas particle resolutions of $M_{\mathrm{DM}}=7.5\times10^{6}M_{\odot}$ and $M_{\mathrm{gas}}=1.4\times10^{6}M_{\odot}$ respectively.
We also use data from the EAGLE suite \cite{McAlpine2016eagle} to augment our sample of simulated systems.
Specifically, we select the fiducial Ref-L0100N1504 run, which -- with an associated simulation box length of $100\,\mathrm{Mpc}$, cosmological parameters from Planck \cite{PlanckCollaboration2014}, as well as comparable dark and gas particle resolutions of $M_{\mathrm{DM}}=9.70\times10^{6}M_{\odot}$ and $M_{\mathrm{gas}}=1.81\times10^{6}M_{\odot}$ respectively -- ensures a general degree of consistency with TNG100-1.

M31-like systems in TNG and EAGLE are obtained as follows.
We first identify M31's mass analogs by searching for dark haloes with virial masses within $M_{200}=[0.5,\,3.0]\times10^{12}M_{\odot}$, a range which accommodates literature estimates of M31's halo mass within $0.7 \sim 2.6\times10^{12}M_{\odot}$ \cite{Tamm2012stellar,Fardal2013inferring,Yuan2022constraining} while also including the more numerous, lower-mass haloes for improved statistics.
This range encompasses but is broader than the mass criteria previously used in studies of M31's satellite plane \cite{Bahl2014comparison,Ibata2014thousand}, and was chosen to increase our final sample of M31 analogs.
Within our adopted mass range, minimum isotropic frequencies (MIFs) within the combined TNG and EAGLE sample does not demonstrate a significant correlation with halo mass.
Kendall's Tau test, a statistical measure of rank correlation, yields coefficients (and corresponding p-values) of $1.1\times10^{-2}$ ($p=0.43$) and $2.4\times10^{-2}$ ($p=0.52$) for cones of maximum asymmetry and companion-locked cones respectively.

Next, we search for companions by identifying haloes with a mass above $0.25\,M_{200}$ within a distance of $5\,R_{200}$ (approximately 1 Mpc), which can be more massive than the M31 analogs themselves.
Fractional thresholds are used instead of absolute values due to the relatively large range covered by our halo mass criterion.
If no such companions exist, we consider the system "isolated".
If exactly one companion is found, we require it to fulfil a so-called reciprocity criterion.
We search for its companions in turn using the same \emph{absolute} mass and distance conditions as initially calculated for the original M31-like halo.
If the companion's only companion is the original halo, the reciprocity criterion is satisfied and we designate the system as "paired".
Otherwise, the original halo is rejected.
We also do not consider triplets in this work, and thus reject M31-like haloes in the case where it has two or more companions to avoid crowded fields.
The distribution of host and companion masses in paired simulated systems along their 3D separations is shown in Supplementary Figure 1.

We assign satellites to a given halo by identifying all subhaloes within a distance between $20\,\mathrm{kpc}$ and $2\,R_{200}$, a range which fully encompasses the radial distribution of the adopted M31 satellites \cite{Savino2022hubble}.
Using the K-band luminosity of the Andromeda dwarfs as an estimator of their stellar mass assuming $M_*/L_K \simeq 1M_{\odot}/L_{\odot}$ \cite{Bell2003optical}, 11 of the 37 M31 satellites have stellar masses below TNG100 and EAGLE's specified $M_{\mathrm{gas}}$.
We do not expect this to significantly affect our results, however, since the phase-space distribution of satellites is generally unaffected by internal stellar feedback.
Simulated subhaloes are ranked in descending order by stellar mass, then dark mass once no star particles are available.
All M31 analogs are required to host at least 37 satellites.
If more are found, the ranked sample is truncated to the most massive 37 satellites in order to match the observational sample.
In addition, we filter out systems currently experiencing major mergers by requiring that no satellite has a total mass greater than a quarter of the host galaxy's mass.
In this way, we obtain a total of 1268 and 1107 systems sampled from the hydrodynamic TNG100-1 and EAGLE Ref-L0100N1504 runs.
The TNG sample consists of 184 paired and 1084 isolated systems -- from EAGLE, we obtain 141 and 966 systems respectively.

We mock-observe each simulated system with Andromeda-like distance uncertainties to yield 100 Monte Carlo realisations per system.
Errors are drawn from a Gaussian distribution from each of the M31 satellites and applied randomly without replacement to each system of simulated satellites, and the same is performed for the host galaxy itself.
For isolated systems, mock-observation is performed along line-of-sight vectors drawn from isotropy.
For paired systems, two sets of realisations are produced: one adopts isotropic mock-observation vectors, while the other uses the line-of-sight between the given M31 analog and its sole companion.
When using metrics of asymmetry (see below section) that do not require a companion galaxy, the former set is used.
If a companion's direction is required by a given metric, the latter realisations are adopted instead.
Since we apply distance uncertainties to the M31 system's simulated analogs, we only consider the expected positions for the Andromeda satellites for the remainder of this work.

\subsection*{M31's well-populated asymmetry using cones of maximum asymmetry}

Minimum isotropic frequencies (see main text) enable us to compare the observed Andromeda system and its simulated analogs using the significance of their asymmetrical satellite distributions.
When disregarding the asymmetry's alignment with any companion galaxies, we obtain an incidence of M31-like lopsided systems of $4.5\%$ and $6.2\%$ in TNG and EAGLE respectively.
As is the case when using companion-locked cones, however, a majority of these simulated analogs are most significant with respect to smaller clusters of satellites (see Supplementary Figure 2). 
Only $0.1\%$ of simulated realisations demonstrate an equal or smaller MIF than the M31 system while also matching the observed asymmetry's population-wide nature with a satellite count of $N_{sat}\geq36$.
The satellite system of Andromeda remains rare in a full cosmological context regardless of its asymmetry's orientation.

\subsection*{Incidence of M31-like systems using alternative metrics}

Supplementary Table 1 lists the incidence of M31-like lopsided systems in cosmological simulations when adopting alternative metrics of asymmetry.
Unlike the cone-based approach which places systems in a two-dimensional $N_{\theta}-\theta$ space, these metrics are one-dimensional and thus do not require any consideration of the look-elsewhere effect.
The Andromeda system's incidence is consistently lower when adopting metrics that take its companion galaxy's direction into account.
The remaining metrics yield a $2-6\%$ incidence depending on the simulation used, a result roughly consistent with the $5\%$ obtained using cones of maximum asymmetry at each simulated analog's minimum isotropic frequency.
The fact that Andromeda's incidence according to its normalized centroid shift -- the only metric that takes satellite distances into account -- is similar or higher than the other scale-free metrics may suggest that the observed lopsidedness is not driven by an offset of M31's position from its satellite distribution.

\subsection*{Robustness with simulation resolution}

Andromeda's analogs in cosmological simulations must have 37 satellite galaxies within $2\,R_{200}$ to match the observed M31 system.
At TNG100 and EAGLE's baryonic resolution, many of the lower-mass satellites among the 37 per system are dark subhaloes devoid of any stellar particles.
Baryonic processes within individual dwarfs are not expected to pose a significant impact on their phase-space distribution at scales of hundreds of kpc \cite{Pawlowski2021phase-space}.
Nevertheless, we test for robustness in the higher-resolution TNG50 run \cite{Pillepich2019first}, which adopts the same cosmology and feedback prescription as TNG100 with 15-fold increase in particle resolution.
Due to TNG50's smaller simulation volume, we only obtain a total of 256 isolated and 54 paired M31 analogs.
According to their minimum isotropic frequencies using cones of maximum asymmetry, $5.7\%$ of simulated realisations are as significantly lopsided as the observed Andromeda system.
None of the paired analogs can reproduce the M31 satellites' asymmetry towards the Milky Way, thus setting an upper limit of $2\%$ ($<1/54$) on the true incidence of Andromeda-like systems in TNG50.
In both cases, the resulting incidences are fully consistent with our results in TNG100 and EAGLE.
The Andromeda system's tension with CDM simulations is not an artefact of an insufficient particle resolution.

\subsection*{Exclusion of outlying satellites}

Of \emph{Savino et al.} \cite{Savino2022hubble}'s published sample of stellar systems, only 2 satellite galaxies -- the Peg DIG and IC 1613 -- lie beyond $1.5\,R_{200}$ of Andromeda.
It has been reported that the lopsidedness of satellite galaxy systems is driven by more distant satellites in both observations \cite{Brainerd2020lopsided} and simulations \cite{Gong2019origin,Wang2021lopsided}.
Hence, we exclude these two outlying galaxies and perform the same analysis as described in the main text using an alternative sample of 35 satellites and a reduced search volume of $1.5\,R_{200}$ around simulated hosts.
Since the Peg DIG and IC 1613 are both located within the excess of satellites towards the Milky Way, their removal does not change the cone opening angles at which the Andromeda system is most significant, which remains at $101^{\circ}$ and $106.5^{\circ}$ for cones of maximum asymmetry and companion-locked cones respectively.
The incidence of M31-like lopsided systems in simulations with respect to the Milky Way's direction decreases slightly to $0.2\%$, while the rarity of the observed system without a preferred direction remains unchanged at $5\%$.
The outlying Peg DIG and IC 1613 satellites are not key drivers of the observed asymmetry, and their exclusion does not significantly affect the lopsided Andromeda system's tension with cosmological expectations.

\subsection*{Robustness with paired host selection criteria}

Our selection criteria for M31 analogs in simulations was chosen to ensure a sufficiently large sample to obtain robust statistics, but the adopted range of parameters may deviate from strictly Local Group-like systems due to the enhanced abundance of lower-mass haloes in the CDM mass function. 
We test whether the main outcomes of this work is sensitive to our choice of selection criteria by subdividing our fiducial sample of paired TNG100 and EAGLE hosts into two bins about the median host mass ($M_{200}=1.33\times10^{12}\,M_{\odot}$), relative companion mass ($M_{\mathrm{comp}}=0.46\,M_{200}$), or separation ($D_{\mathrm{comp}}=4.08\,R_{200}$).

The results of this comparison are shown in Supplementary Table 2.
We focus on the incidence of paired simulated systems that match the M31 system's companion-locked MIF, $f_{\mathrm{MIF}}$, as the main result of this work.
The two bins of varying host mass demonstrate $f_{\mathrm{MIF}}$ consistent within $\Delta f_{\mathrm{mif}}=0.09\%$.
A smaller pair separation and higher companion mass do slightly enhance the incidence of M31-like asymmetric systems, but neither can achieve $f_{\mathrm{MIF}}>0.5\%$.
This increase in incidence is not recovered to the same extent for cones of maximum asymmetry, suggesting that this effect only impacts the direction of asymmetry (but not its magnitude).
We also test whether adopting stricter analogs of the M31 system ($M_{200}=[1,2]\times10^{12}\,M_{\odot}$ and $M_{\mathrm{comp}}=[0.5,1.5]\,M_{200}$) alleviates its tension with cosmological expectations.
This selection in fact reduces $f_{\mathrm{MIF}}$ for both companion-locked cones and cones of maximum asymmetry from our fiducial results, but the low sample size of 54 simulated systems across TNG100 and EAGLE renders the results of this comparison unreliable.
Finally, we check whether simulated M31 analogs with more companions more massive than the respective analog ($N_{\mathrm{sys}}=63$) demonstrate an enhanced degree of asymmetry.
We recover respective incidences of 0.30\% and 5.29\% for companion-locked cones and cones of maximum asymmetry, a result consistent with sample (i) in Supplementary Table 2.
In summary, we find no evidence that adjustments in our selection criteria for simulated M31 analogs would significantly alleviate the reported discrepancy with the observed M31 system.

\subsection*{Impact of survey completeness}

The current census of Andromeda satellites is not complete, and the prominent excess of galaxies on the near side of M31 naturally points to incompleteness as a possible factor.
Doliva-Dolinsky \emph{et al.} \cite{Doliva-Dolinsky2022pandas} calculated detection limits within the PAndAS footprint, reporting that dwarfs brighter than $M_V = -7.5$ should be observable throughout the M31 halo.
Of the 37 M31 satellites in Savino \emph{et al.} \cite{Savino2022hubble}, 30 lie above this luminosity threshold.
5 of the 7 fainter satellites are located on the near side of M31 (And XXVI lies only a few degrees further), and their removal from our working sample is expected to increase the incidence of similarly lopsided systems.

To fully compensate for any possible incompleteness, the ideal approach would be to prepare a sample of simulated Andromeda analogs also with 30 satellites brighter than $M_V = -7.5$.
However, TNG100 only produces luminous dwarfs until $M_V = -9$ due to its limited baryonic resolution, and only 15 such paired systems are available in the higher-resolution TNG50 run.
While none of these 15 systems match the Andromeda system's MIF using companion-locked cones, the resulting maximum incidence of $f_{\mathrm{MIF}}<6.7\%$ is limited in its use due to it being an order of magnitude higher than our fiducial sample's $f_{\mathrm{MIF}}$.
To obtain meaningful results, we instead assume that the phase-space distribution of satellites around their host is independent of their exact luminosity function -- a simplifying assumption also adopted in most studies quantifying the significance of observed satellite planes in simulations \cite{Pawlowski2021co-orbitation,Ibata2014thousand,Muller2021coherent,Samuel2021planes} -- and truncate each simulated analog to its 30 most massive satellites in a manner similar to our fiducial method.
In support of this, we point out that $f_{\mathrm{MIF}}$ does not differ significantly when binning over the fiducial sample's range of host (and thus satellite) mass regimes (see Supplementary Table 2), and the luminosity function of our selected analogs need not be an exact match to the M31 system in order to set a baseline for satellite angular phase-space distributions expected in a $\Lambda$CDM context.

Of 411 paired Andromeda analogs and their realisations, $0.5\%$ match the M31 system's MIF for companion-locked cones -- less than a twofold increase in incidence from the full satellite sample.
This result may be interpreted as the upper bound for the Andromeda system's incidence, such that $0.3-0.5\%$ of simulated analogs demonstrate a equal or more significant asymmetry.
When considering cones of maximum asymmetry instead, we recover a higher $f_{\mathrm{MIF}}$ of $8.6\%$, which represents an increase from the full sample at a similar proportion.
However, 74\% of these asymmetric analog realisations reach their MIF at $\theta<60^{\circ}$ -- thus representing a more concentrated cluster of satellites rather than the observed population-wide asymmetry.
In the 30-satellite sample, only $0.2\%$ of realisations match both the M31 system's MIF and the opening angle at which this is achieved.

Compensating for the impact of survey completeness does increase the incidence of sufficiently asymmetric analogs, but incompleteness is not the primary driver of the observed asymmetrical distribution nor the tension it holds with cosmological expectations.
Indeed, we point out that the 7 satellites that lie below the completeness threshold demonstrates a mean angular separation from the Milky Way-M31 line-of-sight of $70\pm8^{\circ}$, a result fully consistent with that of the "complete" 30-satellite sample at $69\pm9^{\circ}$.
The fainter 7 satellites thus display a degree of asymmetry near-identical to the brighter 30, and we find no evidence that incompleteness artificially enhances the observed lopsidedness.
Furthermore, a detailed analysis of the satellite system using forward modelling \cite{Doliva-Dolinsky2023pandas} rules out the asymmetry (within the PAndAS footprint) as a straightforward consequence of incompleteness alone at $>99.9\%$ confidence.
We also point out that Doliva-Dolinsky \emph{et al.}'s analysis only requires $\sim80\%$ of the M31 satellites to be located on the hemisphere facing the Milky Way.
Explaining the more significant statistic of 36 out of 37 satellites within $106.5^{\circ}$ of the Milky Way's direction using incompleteness alone would be even less feasible.

\renewcommand{\thefigure}{\arabic{figure}}


\newpage

\section*{Data Availability}
This work makes use of publicly available data products from the IllustrisTNG and EAGLE simulation suites, provided by the IllustrisTNG Collaboration and the Virgo Consortium respectively.
The observational data of the M31 satellite system used in this work is published in \emph{Savino et al.} \cite{Savino2022hubble}.
Correspondence and requests for other materials should be addressed to K.J.K at \texttt{kkanehisa@aip.de}.

\section*{Code Availability}

The code used to produce this study is not publicly available but can be communicated in response to reasonable requests.

\section*{Acknowledgments}

We thank O. Müller for interesting discussions and helpful inputs. The authors also thank the anonymous reviewers for their constructive comments and suggestions that have helped us to improve the manuscript.

\paragraph{\textbf{Funding}}
K.J.K. and M.S.P. acknowledge funding via a Leibniz-Junior Research Group (project number J94/2020).

\paragraph{\textbf{Author contributions}}
K.J.K. lead the investigation, performed the formal analysis, and wrote the initial manuscript. M.S.P. supervised the design and findings of this work. N.L. contributed to the interpretation of the results and revisions of the methodology. All authors contributed to the writing of the final manuscript.

\paragraph{\textbf{Competing interests}}
The authors declare no competing interests.

\newpage


\begin{figure}[ht]
    \centering
    \includegraphics[width=0.9\textwidth]{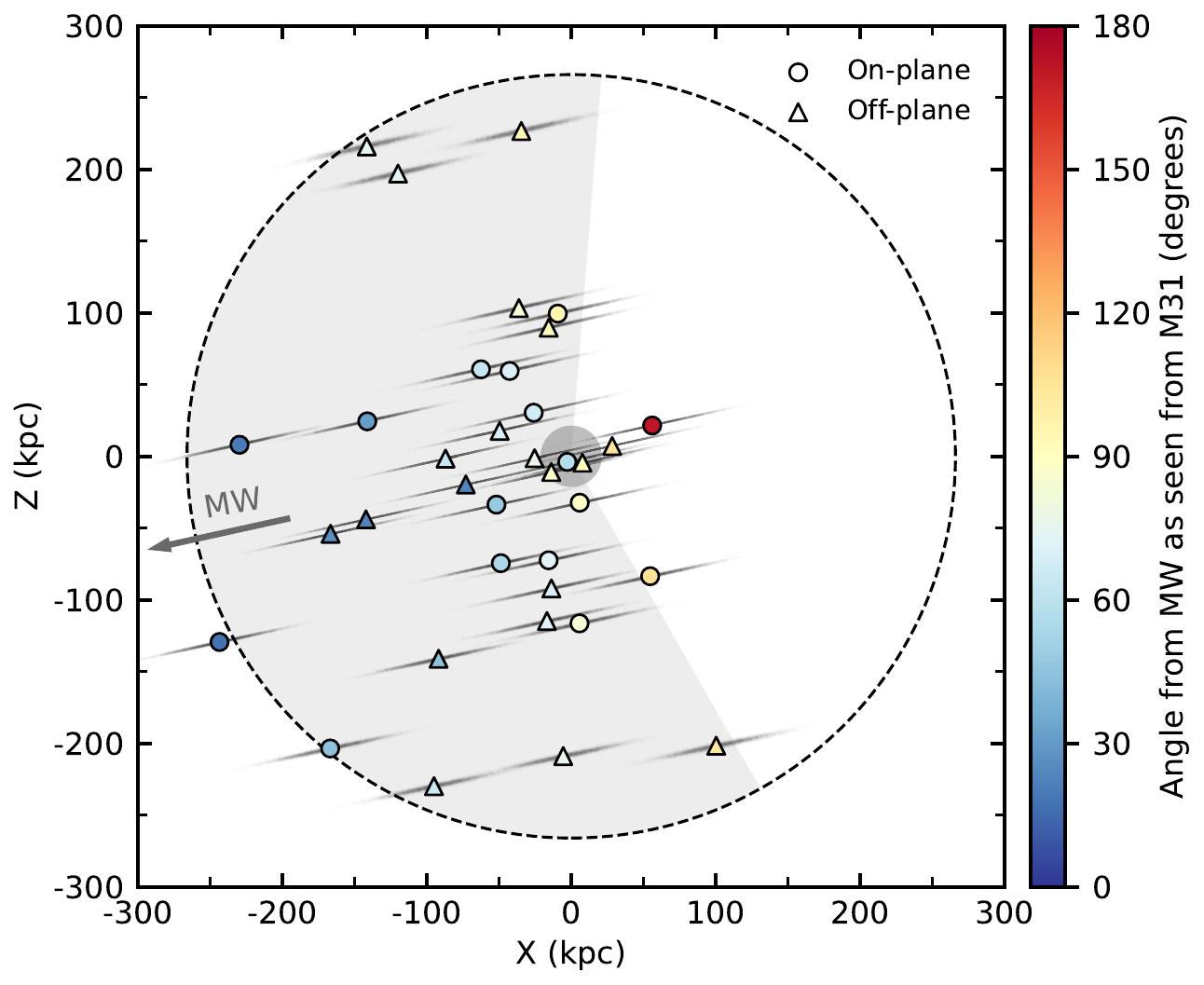}
    \caption{\textbf{A side-on view of Andromeda's asymmetrical satellite distribution.}
    Figure constructed with data from \emph{Savino et al.} \cite{Savino2022hubble}.
    Satellites are projected onto the XZ-plane of our adopted M31-centric coordinate system \cite{McConnachie2006satellite} -- the positive X-axis has the same azimuth ($l=0^{\circ}$) as the Milky Way while the positive Z-axis points to M31's north galactic pole.
    The shaded gray circle represents the position of M31.
    Satellites are coloured by their angular separation from the Milky Way as seen from M31's expected position, and corresponding Monte Carlo uncertainties are drawn in black.
    The dashed circle encompasses M31's virial radius of $266\,\mathrm{kpc}$ \cite{Fardal2013inferring}, while the shaded sector within represents a 2D projection of the angular region within $106.5^{\circ}$ of the Milky Way, in which all but one of M31's 37 satellites are contained.
    }
    \label{fig:1}
\end{figure}

\begin{figure}[ht]
    \centering
    \includegraphics[width=1\textwidth]{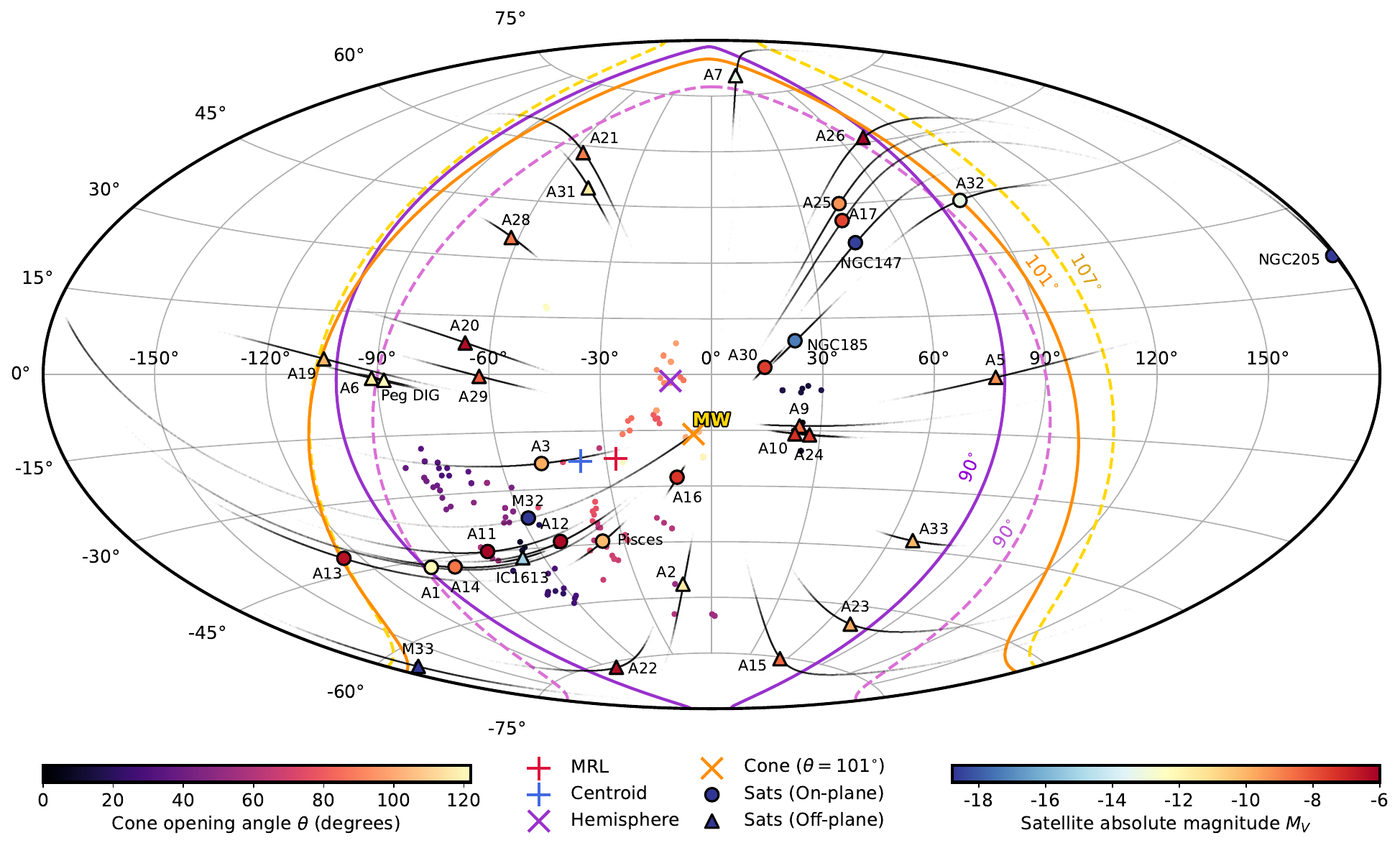}
    \caption{\textbf{Satellite galaxy positions as seen from Andromeda.}
    This equal-area Hammer-Aitoff sky projection in our adopted M31-centric coordinate system \cite{McConnachie2006satellite} shows the expected positions of 37 M31 satellites, coloured according to their absolute magnitudes as compiled in \cite{Savino2022hubble}.
    Corresponding Monte Carlo distance uncertainties are drawn in black.
    Satellites with over a 50\% chance of participating in M31's satellite plane \cite{Savino2022hubble} are drawn as circles, while off-plane satellites are marked with triangles.
    The coloured smaller points each represent the direction of the most populated conical region with a given opening angle $\theta$.
    The purple cross and line indicate the orientation and bounds of the hemispheric region ($\theta=90^{\circ}$) that contains a maximum of 32 satellites, while the dashed purple line delineates a hemispheric region with 29 satellites locked in the direction of the Milky Way.
    The orange cross and line represent the orientation and bounds of the most significant cone ($\theta=101^{\circ}$) containing 36 satellites, while the dashed orange line corresponds to the smallest cone also containing 36 satellites ($\theta=106.5^{\circ}$) but oriented towards our Galaxy.
    The direction of the satellite distribution's geometric centroid and Mean Resultant Length \cite{methods} are shown as blue and red plus symbols.
    Andromeda's satellite galaxies form a strongly asymmetrical distribution that is aligned towards the Milky Way.}
    \label{fig:2}
\end{figure}

\begin{figure}[ht]
    \centering
    \includegraphics[width=1\textwidth]{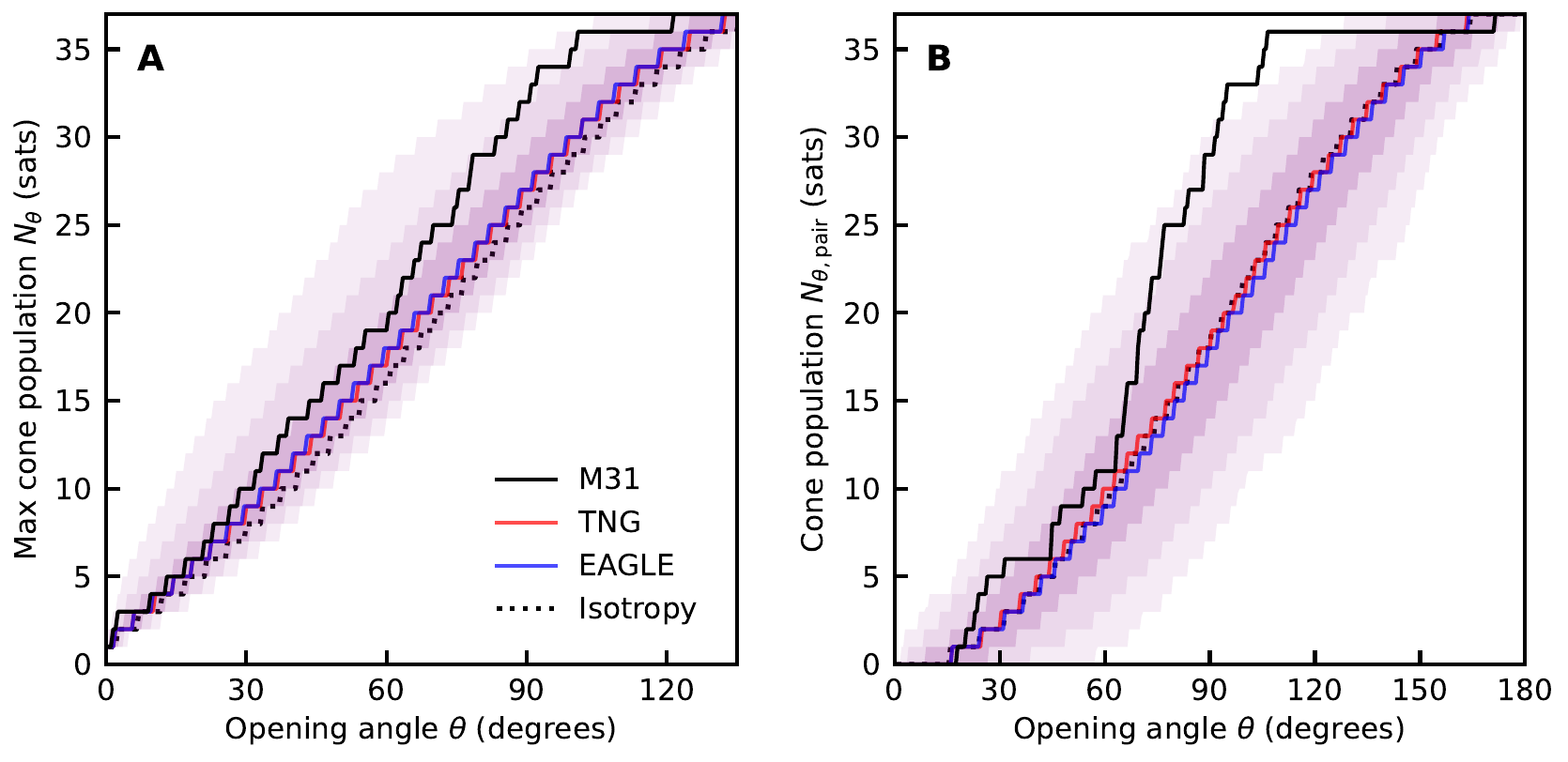}
    \caption{\textbf{Comparison to $\Lambda$CDM simulations.}
    Maximum number of satellites contained within a cone facing in any direction (\textbf{A}) and the number of satellites contained within a cone oriented towards the companion host galaxy (\textbf{B}) with opening angle $\theta$.
    Results for the observed Andromeda satellites are drawn by black lines.
    The median relationship for the M31 system's analogs in the TNG and EAGLE simulations are shown as red and blue lines respectively, while the combined $1\sigma$, $2\sigma$, and $3\sigma$ spreads for all simulated analogs are delineated in purple.
    The isotropic expectation is plotted with grey dotted lines.
    A significant deviation of Andromeda's black line from the simulated medians represents a departure from $\Lambda$CDM expectations.}
    \label{fig:3}
\end{figure}

\begin{figure}[ht]
    \centering
    \includegraphics[width=1\textwidth]{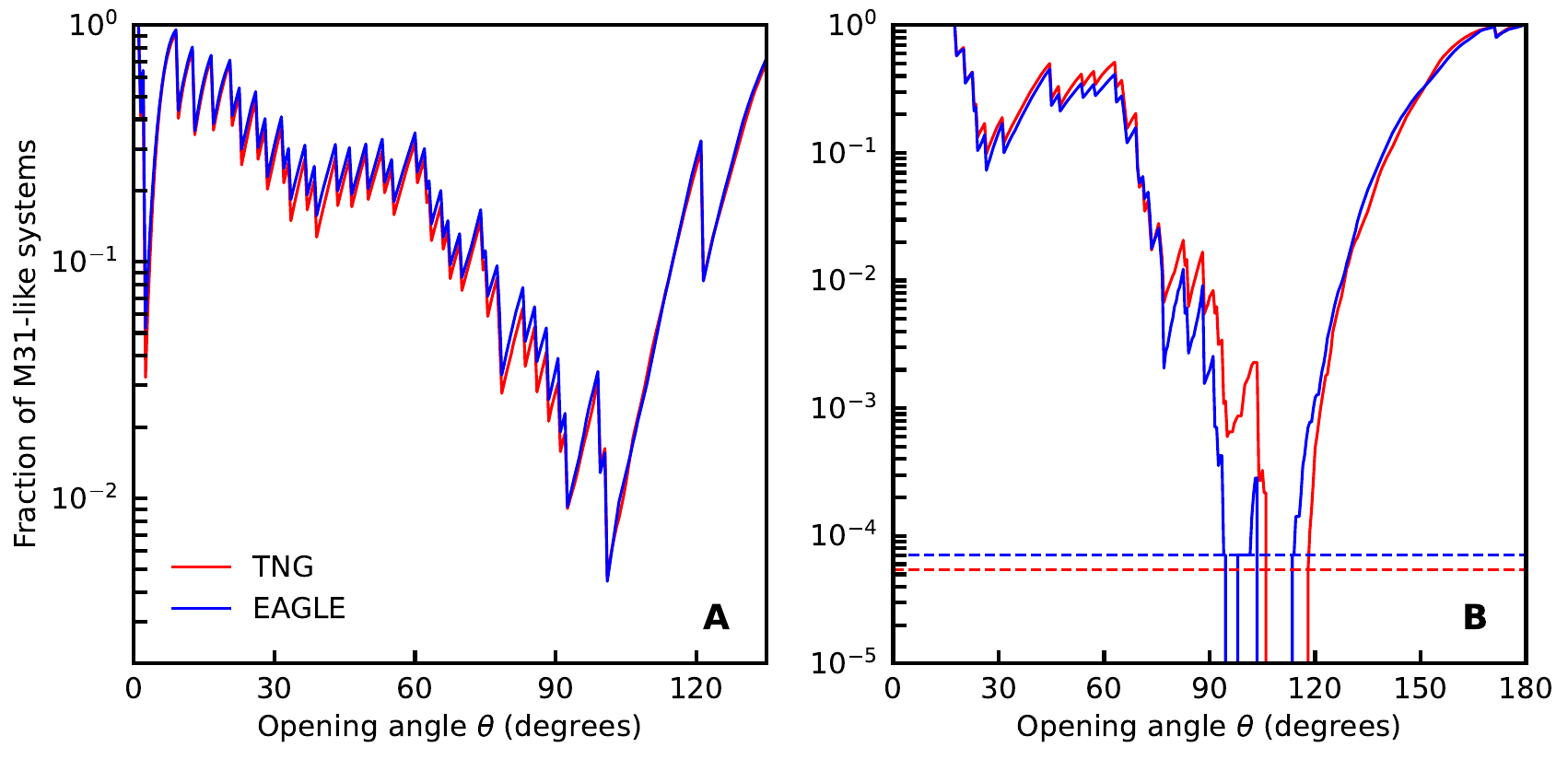}
    \caption{\textbf{Simulated frequency of M31-like lopsided systems.}
    The fraction of simulated M31 analogs that demonstrate an equal or greater degree of asymmetry as the Andromeda system are plotted as a function of the opening angle $\theta$ of cones either \textbf{A}: facing in the direction that maximises their contained satellites, or \textbf{B}: oriented towards their host galaxy's companion.
    Comparisons for analog systems sampled from TNG and EAGLE are drawn as red and blue lines respectively.
    In \textbf{B}, analog fractions corresponding to exactly 1 simulated realisation are indicated by dashed lines of the matching colour.
    Each trough corresponds to an additional satellite within the conical region.
    }
    \label{fig:4}
\end{figure}

\begin{figure}[ht]
    \centering
    \includegraphics[width=0.7\textwidth]{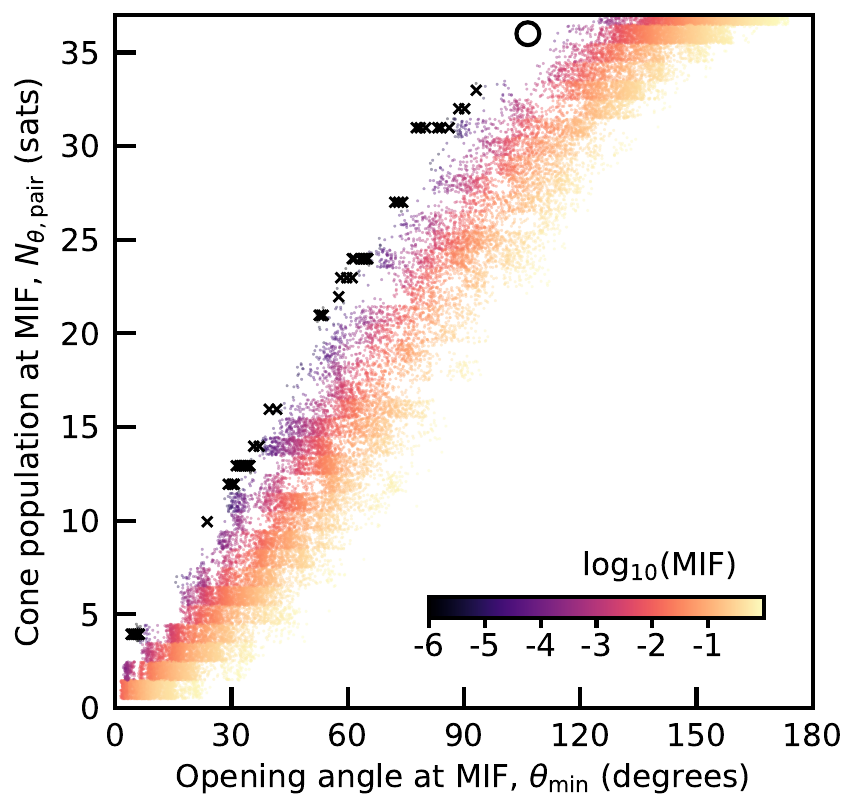}
    \caption{\textbf{The most significant companion-locked cones and their satellite populations.}
    Each simulated analog's cone opening angle at minimum isotropic frequency (MIF) and corresponding satellite population are plotted on the horizontal and vertical axes respectively, when cones are oriented towards the analog's companion galaxy.
    Most systems are coloured according to their MIF, while the few simulated analogs that demonstrate MIFs equal or less than that of Andromeda are drawn as black crosses.
    Satellite counts for coloured systems are shifted within $[-0.5,0.5]$ for visibility.
    The result for the M31 system itself is indicated by the white circle.
    The Andromeda system's asymmetry towards the Milky Way is more population-wide than any of its similarly significantly lopsided analogs.}
    \label{fig:5}
\end{figure}

\renewcommand{\thefigure}{S\arabic{figure}}
\setcounter{figure}{0}

\begin{figure}
    \centering
    \includegraphics[width=0.8\textwidth]{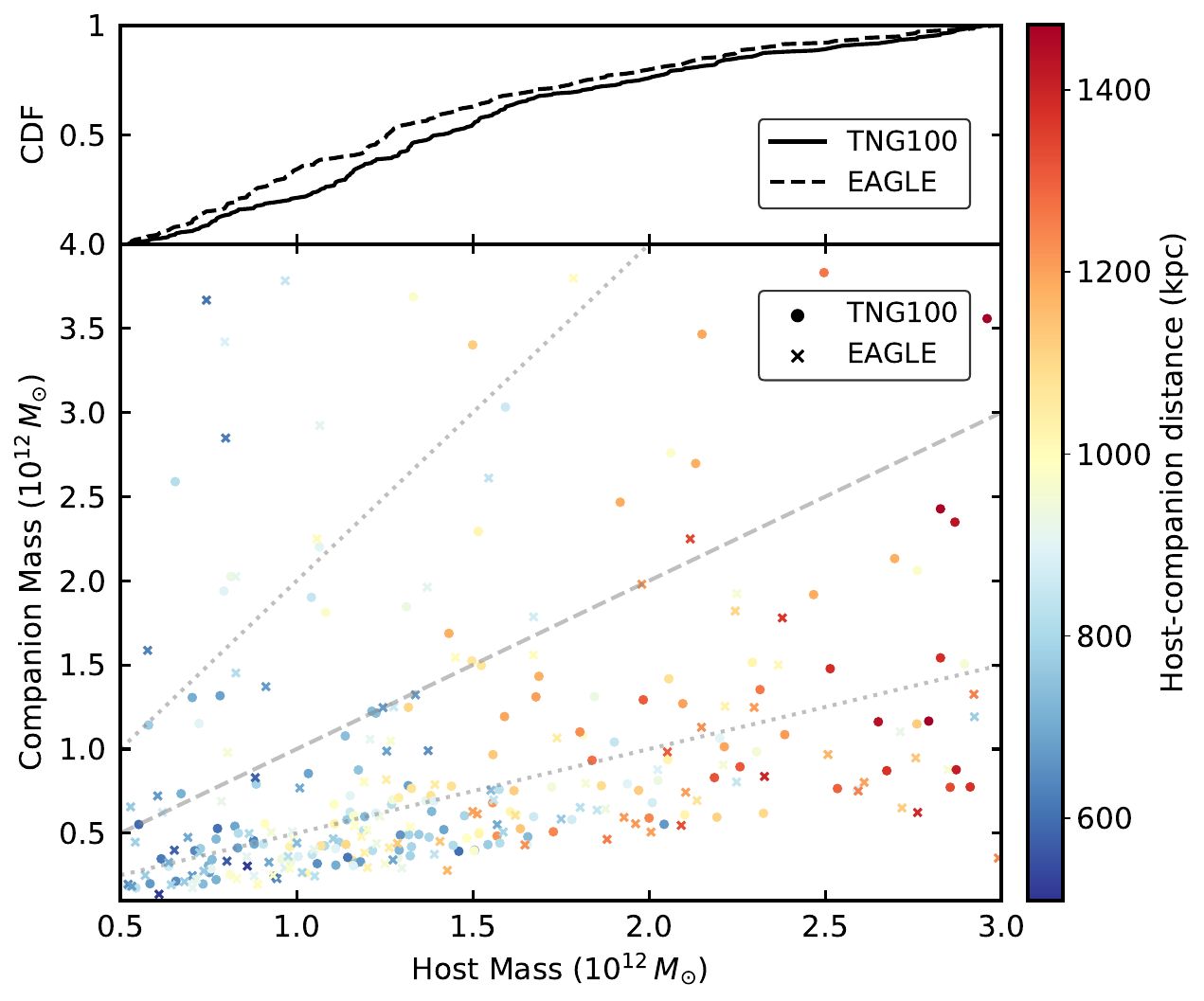}
    \caption{\textbf{Masses and separations of paired hosts in simulations.}
    The halo masses of paired hosts and their companions in TNG100 and EAGLE are coloured by the distance between them.
    The grey dashed line represents the 1:1 mass ratio threshold between hosts and companions, while the lower and upper dotted lines indicate where the host is double the companion's mass and vice versa.
    The upper CDF plots the distribution of host masses in TNG100 and EAGLE.
    A majority of the paired analogs sampled cover a broader range in parameter space than observational estimates but are generally consistent with the M31 system.
    }
    \label{fig:hosts}
\end{figure}

\begin{figure}
    \centering
    \includegraphics[width=0.7\textwidth]{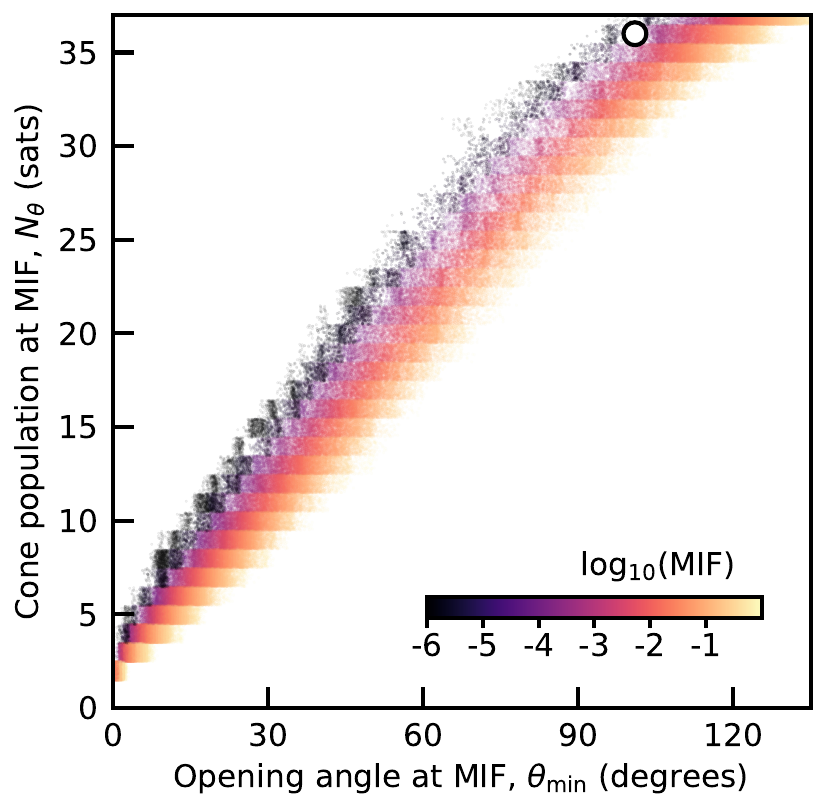}
    \caption{\textbf{The most significant cones of maximum asymmetry and their satellite populations.}
    Each simulated analog's cone opening angle at minimum isotropic frequency (MIF) and corresponding satellite population are plotted on the horizontal and vertical axes respectively.
    Satellite counts are shifted within $[-0.5,0.5]$ for visibility.
    Most systems are coloured according to their MIF, while simulated analogs that demonstrate MIFs equal or smaller than that of Andromeda are drawn in black.
    The result for the M31 system itself is indicated by the white circle.
    Andromeda's asymmetry remains more population-wide than most of its similarly lopsided analogs, even when disregarding its close alignment with the Milky Way.
    }
    \label{fig:minisofreq_free}
\end{figure}

\clearpage

\begin{table*}
	\centering
	\caption{\textbf{Incidence of M31-like lopsided systems in simulations using alternative metrics.} From the left, metrics are: the maximum number of satellites in a given hemisphere, the number of satellites contained within the hemisphere facing a companion galaxy, the normalized shift between the host galaxy's position and its satellites' geometric centroid, the Mean Resultant Length, the mean of all satellite pairwise angles, and the mean of all satellite angular offsets from a companion galaxy's direction.}
	\vspace{2mm}
	\begin{tabular}{ll|ll|l|l|ll}
	    \hline
        \multirow{2}{*}{Run} & \multirow{2}{*}{Sample} & \multicolumn{2}{c|}{Hemisphere $(\%)$} & Centroid $(\%)$ & MRL $(\%)$ & \multicolumn{2}{c}{Pairwise $(\%)$} \\
        & & $N_{\mathrm{lop}}$ & $N_{\mathrm{lop,pair}}$ & $d_{\mathrm{norm}}$ & $\bar{R}$ & $\theta_{\mathrm{lop}}$ & $\theta_{\mathrm{lop,pair}}$ \\
        \hline
        \multirow{2}{*}{TNG100} & Isol & 2.7 & - & 4.6 & 4.2 & 4.2 & - \\
         & Pair & 4.1 & 0.7 & 4.5 & 4.8 & 4.7 & 1.3 \\
        \multirow{2}{*}{TNG50} & Isol & 3.8 & - & 5.8 & 5.3 & 5.4 & - \\
         & Pair & 1.7 & 0.4 & 3.2 & 4.2 & 4.4 & 2.6 \\
        \multirow{2}{*}{EAGLE} & Isol & 3.6 & - & 5.9 & 6.0 & 6.1 & - \\
         & Pair & 2.5 & 0.2 & 5.9 & 5.5 & 6.0 & 0.8 \\
		\hline
	\end{tabular}
	\label{tab:dirfreq_1d}
\end{table*}

\begin{table*}
	\centering
	\caption{\textbf{Incidence of paired M31-like lopsided systems using minimum isotropic frequencies.} For both companion-locked cones and cones of maximum asymmetry, shown are: $f_{\mathrm{MIF}}$ (the fraction of simulated realisations that match the M31 system's MIF at any $\theta$), $N_{\mathrm{system}}$ (the simulated sample size), and $f_{\mathrm{true}}$ (the fraction of realisations that match the M31 system's MIF at an opening angle equal or greater than that most significant for the observed system itself). All samples are based on the fiducial sample (a) adopted in this manuscript with cuts made in satellite search radius (b), satellite count based on completeness in M31's virial volume (c), host mass (d,e), pair separation (f,g), or companion mass (h,i,j). Sample (k) requires both a halo mass $M_{200}=[1,2]\times10^{12}\,M_{\odot}$ and a companion mass within $[0.5,1.5]\,M_{200}$. Note that $f_{\mathrm{MIF}}$ and $f_{\mathrm{true}}$ may not be clean multiples of $1/N_{\mathrm{sys}}$ due to the Monte Carlo sampling producing 100 realisations per system. The significance of the asymmetric M31 system is generally robust to our criteria for analog selection.}
	\vspace{2mm}
	\begin{tabular}{c|ll|ll|ll}
	    \hline
        & \multicolumn{2}{c|}{Sample} & \multicolumn{2}{c|}{Companion-locked} & \multicolumn{2}{|c}{Maximum asymmetry} \\
        & Name & $N_{\mathrm{sys}}$ & $f_{\mathrm{MIF}}\,(\%)$ & $f_{\mathrm{true}}\,(\%)$ & $f_{\mathrm{MIF}}\,(\%)$ & $f_{\mathrm{true}}\,(\%)$ \\
        \hline
        (a) & Fiducial ($N=37$) & 325 & 0.28 & 0 & 5.6 & 0.11 \\
        (b) & $<1.5R_{200}$ ($N=35$) & 325 & 0.12 & 0 & 5.7 & 0.05 \\
        (c) & Complete ($N=30$) & 411 & 0.50 & 0.03 & 8.6 & 0.17 \\
        (d) & $M_{200} < 1.33\times10^{12}\,M_{\odot}$ & 163 & 0.33 & 0 & 5.8 & 0.09 \\
        (e) & $M_{200} \geq 1.33\times10^{12}\,M_{\odot}$ & 162 & 0.24 & 0 & 5.5 & 0.12 \\
        (f) & $D_{\mathrm{pair}} < 4.08\,R_{200}$ & 161 & 0.44 & 0 & 6.2 & 0.12 \\
        (g) & $D_{\mathrm{pair}} \geq 4.08\,R_{200}$ & 164 & 0.12 & 0 & 5.1 & 0.09 \\
        (h) & $M_{\mathrm{comp}} < 0.46\,M_{200}$ & 164 & 0.21 & 0 & 5.5 & 0 \\
        (i) & $M_{\mathrm{comp}} \geq 0.46\,M_{200}$ & 161 & 0.35 & 0 & 5.8 & 0.22 \\
        (j) & $M_{\mathrm{comp}} \geq 1.00\,M_{200}$ & 63 & 0.30 & 0 & 5.3 & 0.10 \\
        (k) & Most Analogous & 54 & 0.02 & 0 & 4.0 & 0.37 \\
	\end{tabular}
	\label{tab:mif_incidence}
\end{table*}

\clearpage

\bibliography{manuscript}
\bibliographystyle{Science}

\end{document}